\documentstyle[12pt,epsfig]{article}
\newlength{\dinwidth}
\newlength{\dinmargin}
\newcommand{\ba}{\begin{array}}
\newcommand{\ea}{\end{array}}
\newcommand{\beq}{\begin{equation}}
\newcommand{\eeq}{\end{equation}}
\newcommand{\bea}{\begin{eqnarray}}
\newcommand{\eea}{\end{eqnarray}}

\def\bce{\begin{center}}
\def\ece{\end{center}}

\def\nonu{\nonumber}

\def\pa{\partial}
\def\al{\alpha}
\def\be{\beta}

\def\de{\delta}

\def\ep{\epsilon}

\def\la{\lambda}

\def\si{\sigma}

\def\eps6{{\displaystyle \mathop{\epsilon}^{6}}{}}

\def\nab6{{\displaystyle \mathop{\nabla}^{6}}{}}

\def\ba{\begin{array}}
\def\ea{\end{array}}
\def\beq{\begin{equation}}
\def\eeq{\end{equation}}
\def\be{\begin{equation}}
\def\ee{\end{equation}}

\def\la{\lambda}
\def\eps{\epsilon}

\def\ba{\begin{array}}
\def\ea{\end{array}}
\def\beq{\begin{equation}}
\def\eeq{\end{equation}}
\def\be{\begin{equation}}
\def\ee{\end{equation}}

\def\la{\lambda}
\def\eps{\epsilon}

\newcommand{\bean}{\begin{eqnarray*}}
\newcommand{\eean}{\end{eqnarray*}}

\begin{document}
\thispagestyle{empty} \addtocounter{page}{-1}
\begin{flushright}
{\tt hep-th/0409195}\\
\end{flushright}

\vspace*{1.3cm} 
\centerline{ \Large \bf ${\cal N}=1$ Conformal Supergravity 
and Twistor-String Theory }
\vspace*{1.5cm}
\centerline{{\bf Changhyun Ahn}}
\vspace*{1.0cm} 
 \centerline{\it   School of Natural Sciences,
Institute for Advanced Study,
Olden Lane, Princeton NJ 08540, USA}
\centerline{\it  Department of Physics,
Kyungpook National University, Taegu 702-701, Korea}
\vspace*{0.8cm}
\centerline{\tt
ahn@ias.edu 
}  
\vskip2cm

\centerline{\bf Abstract}
\vspace*{0.5cm}

The physical states of ${\cal N}=4$ conformal supergravity  
in four dimensions occur in twistor-string theory by 
Berkovits and Witten(hep-th/0406051).
We study two alternative versions of twistor-string theory
based on the B-model of weighted projective space ${\bf WCP }^{3|2}$
and based on a certain construction involving open strings.
The spacetime fields described by the twistor 
superfields contain the physical states of ${\cal N}=1$ conformal
supergravity from above ${\cal N}=4$ superspace approach.

\baselineskip=18pt
\newpage
\renewcommand{\theequation}{\arabic{section}\mbox{.}\arabic{equation}}

\section{Introduction}
\setcounter{equation}{0}

\indent

Contrary to the fact that the usual string theories are not 
conformally invariant in
the target space(providing Einstein supergravity), 
in twistor-string theory \cite{Witten}  due to 
the superconformal invariance
in spacetime the conformal supergravity arises. 
In four dimensions, the action has a quadratic Weyl tensor.
See \cite{FT} for the review of
conformal supergravity.  

In \cite{BW},
one of the main results was that the gauge-singlet sector 
of twistor-string theory has the same physical states as 
${\cal N}=4$ conformal
supergravity in four dimensions.
The ${\cal N}=4$ chiral superfield which is a function of a chiral 
Minkowski superspace with coordinates coincides with 
a coupling of the abelian gauge 
$g$-field to the boundary of an open string
world sheet.
This boundary corresponds to a point in Minkowski spacetime and
is defined via the twistor equations \cite{Witten}.
To determine the helicities of massless state in Minkowski 
spacetime, the relation \cite{Pen,Ati} 
between the degree of homogeneous 
coordinates of twistor space and the helicity as well as 
the gauge invariance and constraint were used.
In the conformal supergravity side, 
the analysis for the solutions to the field equations
which have higher derivatives(the spin 
$\frac{3}{2}$ field has third-order derivatives and the spin 
$2$ field has fourth-order derivatives) was used.
The occurrence of two states with the same helicity 
where they form a doublet
reflects the nonunitarity of the theory.

In this paper, 
by simply replacing ${\bf CP}^{3|4}$ with 
the weighted projective space ${\bf WCP}^{3|2}$
introduced in \cite{Witten},
we study how their physical states correspond to each other.  
In the twistor-string theory side, this weighted projective 
space has different weights in the fermionic coordinates
and its bosonic submanifold is the same as ${\bf CP}^3$ of
${\bf CP}^{3|4}$.
The only bosonic and fermionic homogeneous 
coordinates of weight one participate in ${\cal N}=1$ 
supersymmetry. 
In the ${\cal N} \leq 4$ conformal supergravity side, 
the number of 
spinor index of chiral field strength superfield depends on the 
number of supersymmetry ${\cal N}$ via $2s= 4-{\cal N}$.
When ${\cal N} \neq 4$, there exists a chiral superfield which
has nonzero spinor indices. 
In particular, when ${\cal N}=1$, the theory can be described by
a chiral superfield(which is a function of 
a chiral Minkowski superspace) with three spinor indices.
By realizing that this ${\cal N}=1$ chiral superfield can be 
obtained from above ${\cal N}=4$ chiral superfield(by taking three
superspace derivatives that do not play a role in ${\cal N}=1$ 
superspace and putting the other three fermionic 
coordinates to zero), 
we study the relation between 
the spacetime fields described by the twistor 
superfields and the physical states of ${\cal N}=1$ conformal
supergravity from the ${\cal N}=4$ superspace approach 
\cite{BW}. 

In section 2, 
we describe 
the twistor-string vertex operators  
in two alternative versions. 
In section 3, we discuss the spectrum of massless fields 
in Minkowski spacetime characterized by twistor fields.
In section 4, we consider the linearized spectrum of
conformal supergravity and compare it with the twistor-string
theory results of previous section.  
In doing this, the role of ${\cal N}=1$ chiral superfield in
${\cal N}=4$ superspace is crucial. 
In section 5, 
we make some remarks after summarizing the main results of 
this paper.

\section{Vertex operator of the B-model on 
${\bf WCP }^{3|2}$}
\setcounter{equation}{0}

\indent

Let us describe the twistor-string vertex operators in 
the open string version first and then we consider  
the topological B-model of 
${\bf WCP }^{3|2}(1,1,1,1|1,3)$ which is denoted by 
${\bf WCP}^{3|2}$ in this paper, for simplicity.

In the open twistorial string theory \cite{Berk},
the worldsheet action can be written as
the homogeneous coordinates(or super-twistor variables) 
\bea
Z^I= (\la^a, \mu^{\dot{a}},\psi^A), \qquad a, \dot{a}, A=1,2
\nonu
\eea
of   
${\bf WCP }^{3|2}$ and its complex conjugates 
$\overline{Z}^I$,
conjugate super-twistor variables $Y_I$ and 
its complex conjugates $\overline{Y}_I$,
a worldsheet $GL(1)$ connection through the covariant derivatives 
plus the action for current algebra.
The conformal dimensions of $(Z^I, \overline{Z}^I, Y_I, 
\overline{Y}_I)$
are $(0,0),(0,0),(1,0)$, and $(0,1)$ respectively.
The physical states are described by the dimension one 
vertex operators that are $GL(1)$-neutral and primary fields
under the Virasoro and $GL(1)$ generators.
The boundary condition for open string 
implies that open string vertex operators
can be written in terms of $Z^I$ and $Y_I$ and current algebra 
variables(not $\overline{Z}^I$ and $\overline{Y}_I$).  

The simplest dimension zero primary fields are 
any function $\phi(Z^I)$(that is $GL(1)$-neutral, that is, 
invariant under $GL(1)$)  
on ${\bf WCP }^{3|2}$. By combining this with any 
current $j_r$ where $r=1,2, \cdots, \mbox{dim} G$ of the current
algebra associated with some group $G$, the Yang-Mills vertex
operator of dimension 1 is given by $j_r \phi^r(Z)$ \cite{Berk}.

On the other hand, 
the conformal supergravity multiplet can be described by
dimension 1 vertex operator by using either $Y_I$ or 
$\pa Z^I$(recall that the
dimension of these is 1). Then one can construct them as follows:   
\bea
Y_I f^I(Z), \qquad g_I(Z) \pa Z^I.
\nonu
\eea 
These are $GL(1)$-invariant if 
$f^I$ carries $GL(1)$ charge 1 for the indices $a, \dot{a}, A=1$
and 3 for the index $A=2$. In other words, 
under the $Z^I \rightarrow t Z^I$ where $Z^I$ stands for 
the homogeneous bosonic and fermionic coordinates of weight 1, 
the $f^I$ scales as $f^I 
\rightarrow t f^I$.
The corresponding $Y_I$ has $GL(1)$ charge $-1$.
Under the $Z^I \rightarrow t^3 Z^I$ 
where $Z^I$ stands for the homogeneous fermionic coordinate
of weight 3, the $f^I$ scales as 
\bea
f^{A=2} 
\rightarrow t^3 f^{A=2}.
\nonu
\eea
The corresponding $Y_I$ has $GL(1)$ charge $-3$.

Similarly the vertex operator corresponding to 
$g_I$ is $GL(1)$-invariant
if $g_I$ carries $GL(1)$ charge $-1$ 
for the indices $a, \dot{a}, A=1$
and $-3$ for the index $A=2$. In other words, 
under the $Z^I \rightarrow t Z^I$ where $Z^I$ stands for 
the homogeneous bosonic and fermionic coordinates of weight 1, 
the $g_I$ scales as $g_I 
\rightarrow t^{-1} g_I$.
Under the $Z^I \rightarrow t^3 Z^I$ 
where $Z^I$ stands for the homogeneous fermionic coordinate
of weight 3, the $g_I$ scales as 
\bea
g_{A=2} 
\rightarrow t^{-3} g_{A=2}.
\nonu
\eea
The constraint $\pa_I f^I=0$ coming from the 
condition of primary field on the vertex operator implies that
the $f^I$ is a volume-preserving vector field.
The other constraint $g_I Z^I=0$ implies that 
$g_I dZ^I$ is well-defined one form.

Let us consider the second version of twistor-string theory 
and 
we denote the complex homogeneous coordinates
of ${\bf WCP }^{3|2}$ as 
$Z^I= (\la^a, \mu^{\dot{a}},\psi^A)$
where $\la^a$ and $\mu^{\dot{a}}$ are bosonic 
spinors of opposite helicity and $\psi^A$ are fermions.
Let us consider the conformal supergravity sector.
In the B-model, closed string modes 
describe the nontrivial deformations of 
the complex structure  of the 
target space ${\bf WCP }^{\prime 3|2}$ which denotes the region
in ${\bf WCP }^{3|2}$ in which the $\la^a$ are not both zero. 
The deformation 
should preserve the holomorphic volume-form $\Omega$.
We cover ${\bf WCP }^{\prime 3|2}$ with open sets $U_i$ 
\cite{PW}.
One can take two open sets, a set $U_1$ characterized by 
$\la^1 \neq 0$ and a set $U_2$ characterized 
by $\la^2 \neq 0$(Note that ${\bf WCP}^{3|2}$ is 
not deformable and rigid). 
On their intersections $U_{ij}$ one glues the open sets $U_i$
through diffeomorphism of the form 
$Z^I \rightarrow Z^I + \epsilon f_{ij}^I$ where $\epsilon$
is an infinitesimal parameter and $f_{ji}=-f_{ij}$.  

One can proceed the discussion of \cite{BW} for our weighted 
projective space and, for example,
the nonlinear action can be written similarly as \cite{BW} and 
is given by
\bea
\int_{{\bf WCP }^{3|2}} d \overline{X}^{\overline{I}}
d \overline{X}^{\overline{J}}
d \overline{X}^{\overline{K}}
 b_{\overline{I} I} N_{\overline{J} \overline{K}}^{I} 
\Omega 
\nonu
\eea
where $X^I$ are local complex coordinates on ${\bf WCP}^{3|2}$
and the complex conjugates $\overline{X}^I$ are bosonic. 
The $\Omega$ is 
holomorphic volume-form or measure. 
From an almost complex structure, one can construct an invariant 
tensor called Nijenhuis tensor $N$.
In Minkowski spacetime, the condition for the vanishing 
of Nijenhuis tensor(coming from the equation of motion $b$) 
corresponds to $F_{abcd}=0$ 
which is the self-dual part of Weyl tensor
and is symmetric in all their indices, according to 
the result of Penrose \cite{Penrose}. 

Then the whole contribution of 
spacetime description of above action  
plus a $G^2$ interaction($G$ is a spin 2-field and part of the
spacetime description of the twistor field $b$) \cite{Witten} 
gives rise to   
$
\int d^4 x \sqrt{g} \left( G^{abcd} F_{abcd} -\frac{1}{2} \ep G^2
\right)
$
which is equal to 
$
\frac{1}{2\ep} \int d^4 x \sqrt{g} F^{abcd} F_{abcd}
$
after integrating out $G$.
A degree one instanton is a copy of 
${\bf CP}^1$ embedded in ${\bf WCP}^{3|2}$ and in the Penrose 
transform, this corresponds to a point in a chiral Minkowski 
superspace. One can define a function by an integral of $b$ over
the curve with moduli $x$ and $\theta_1$. Then   
the $\theta_1$-expansion of this function 
${\cal W}_{abc}^{{\cal N}=1}(x,\theta_1)$ 
is given by
$
{\cal W}_{abc}^{{\cal N}=1}(x,\theta_1) = 
\Phi_{abc}(x) +  
\theta^{d}_1 G_{abcd}(x) + 
\theta_1^d \theta_{1d} 
\Delta_{abc}(x)
$
where the expression for
the effective action quadratic in $b$ over 
the moduli space with an appropriate measure reads
$
\int d^4 x^{a\dot{a}} d^2 \theta^{e}_1 \left[
{\cal W}^{{\cal N}=1}_{bcd}(x,\theta_1) \right]^2 
$.
By substituting ${\cal W}_{abc}^{{\cal N}=1}(x,\theta_1)$  
into this action,
the above $G^2$ interaction can be obtained.

\section{Spectrum of massless fields in Minkowski spacetime}
\setcounter{equation}{0}

\indent

Let us consider the twistor field $f^I(Z)$
which is a function of four bosonic and two fermionic variables 
$\la^a, \mu^{\dot{a}} (a, \dot{a} =1,2)$ and $\psi^A (A=1,2)$. 
Let us denote $\psi^{A=1}$ as $\psi$ and $\psi^{A=2}$ as 
$\chi$
\bea
\psi^{A=1} \equiv \psi, \qquad \psi^{A=2} \equiv \chi. 
\nonu
\eea
Since 
the $f^I(Z)$, for each $I$ except an index $A=2$, 
is homogeneous in $Z^I$
of degree 1, there exist four bosonic and one fermionic helicity
states, each of helicity $3/2$(since a massless state in Minkowski 
spacetime has a helicity $1 +\frac{1}{2} \mbox{deg.}$ 
\cite{Pen,Ati}) when
we put $\psi^A=0$ and do not take the spinor index.
For $A=2$, the $f^{A=2}(Z)$ is homogeneous 
in $Z^I$ of degree 3 and there is one fermionic helicity 
state $5/2$ by same reason.
Both spinor index $a$ and $\dot{a}$ give 
a helicity $\frac{1}{2}$ and $-\frac{1}{2}$. 
Then this leads to two bosonic 
states of helicity 2 and two of helicity 1. 
Moreover there are
two states of helicity $\frac{3}{2}$ and $\frac{5}{2}$. 
According to the arguments of 
\cite{BW}, after taking account of the gauge invariance and the 
constraint, there are two bosonic states of helicity 2(two 
bosonic states of helicity 1 are removed) and two
fermionic states  of helicity $\frac{3}{2}$ and $\frac{5}{2}$.  

Let us recall \cite{Witten} that the action of $SU(2,2|1)$
on the the homogeneous coordinates $(Z^I, \psi)$ is generated by
$5\times 5$ matrices which are supertraceless.
The bosonic conformal algebra 
$SU(2,2)$ which is the covering group 
for $SO(4,2)$ is represented by   
the $15(=4^2-1=\frac{6\times5}{2})$ 
matrices which live in the $4 \times 4$
left-upper part.
A bosonic generator for chiral $U(1)_R$ transformation
is represented by a diagonal matrix with entries 
$\mbox{diag}(1,1,1,1,4)$ up to an overall scale. Moreover 
a spinorial(supersymmetry) generator and a special conformal
supersymmetry generator are realized as the matrices with 
fifth row and fifth column. All $5\times 5$ matrices are 
supertraceless where the trace of  the $4 \times 4$
left-upper part is equal to the $(5,5)$ matrix element.       

Now for nonzero $\psi^A$, one can expand $f^I(Z)$ in powers of
$\psi$ and $\chi$:
\bea
f^I(\la,\mu,\psi,\chi)=
f_0^I(\la,\mu) +f_{1 \psi}^{I}(\la,\mu) \psi +
f_{1 \chi}^{I}(\la,\mu) \chi +
f_{2 \psi \chi}^{I}(\la,\mu) \psi \chi
\nonu 
\eea 
where $f^I_{1\psi}$ is homogeneous
in $(\la, \mu)$ with degree $0$ and provides a massless state
of helicity $1$ 
when we ignore the angular momentum carried 
by the index $I$, as observed above, $f^I_{1\chi}$ is homogeneous
in $(\la, \mu)$ with degree $-2$(leading to a helicity $0$) and 
 $f^I_{2\psi\chi}$ is homogeneous
in $(\la, \mu)$ with degree $-3$(giving a helicity $-\frac{1}{2}$). 
Here the index $I$
denotes by $a, \dot{a}$ or $A=1$. 

Finally,
we list the full structure 
of helicity states described by the field $f^I(Z)$
by taking into account of angular momentum, the gauge invariance and
the constraint
\bea
\la^a f_a(Z) &:& (2, 0), \;\; (\frac{3}{2}, -1), \;\;
\nonu \\
\mu^{\dot{a}} f_{\dot{a}}(Z) &:& (2, 0), \;\; (\frac{3}{2}, -1), 
\;\; 
\label{Ftwistor} \\
f^{A=1}(Z) &:& (\frac{3}{2}, 1), \;\; 
(1, 0), \;\;(0, 0), \;\; (-\frac{1}{2}, 
-1) 
\nonu 
\eea
where the first element is the helicity and the second element
is the $U(1)_R$ representation.
The $U(1)_R$ transformation laws of the fields are determined by 
the chiral weight \cite{FT}.
Compared with the spectrum of ${\cal N}=4$ field contents,
the first two entries in each $f^I(Z)$ are exactly a truncation of 
${\cal N}=4$ spectrum. They have the same helicity and 
$U(1)_R$ charge. 
The state $(2,{\bf 1})$ of ${\cal N}=4$ twistor field 
has $U(1)_R$ charge $0$ and 
the state $(\frac{3}{2},{\bf \overline{4}})$ 
of ${\cal N}=4$ twistor field 
has $U(1)_R$ charge $-1$ and so on.
Since the helicity $\frac{5}{2}$ of  
fermionic state characterized by 
$f^{A=2}(Z)$ is greater than 2, this field is not allowed in 
the conformal supergravity 
\footnote{The possibility of having a spin $s> 2$ 
conformal supergravity
  field
has been discussed in the existence of some nontrivial ${\cal N}>4$
conformal supergravity \cite{dF}. If this theory can be 
coupled to Poincare
supergravity, then these higher spin fields occur in the conformal
supergravity action like $\overline{\zeta}_{\frac{5}{2}} 
\gamma \cdot \pa 
 \Box^2 \zeta_{\frac{5}{2}}$. It was difficult to form
massive higher multiplets in the spectra of the combined conformal
supergravity
and Poincare supergravity system. Even if ${\cal N}>4$ conformal
supergravity
with higher spin fields can exist, they cannot be coupled to the 
corresponding Poincare supergravity consistently. The appearance of
scalar $W_{ABCD}$ \cite{Siegel} 
for ${\cal N} > 4$ restricts some constraint on this
and corresponding conformally invariant action which is not known 
differs from the one in chiral superspace 
because this scalar is not chiral. It is not clear how to construct
the `full' action including higher spins with higher power of
derivatives.}.  
Note that $f^{I}(Z)$ where $I=a$ or $I=\dot{a}$ has no dependence on 
$\chi$ of weight 3 in
${\bf WCP }^{3|2}(1,1,1,1|1,3)$
and is a `truncation' of the ${\cal N}=4$ case
and $f^{A=1}(Z)$ has a dependence on $\chi$. 
Remember that for ${\cal N}=4$ theory, 
all the $\psi^A(A=1,2,3,4)$ play the role of 
the fermionic coordinates in 
the chiral Minkowski superspace through twistor equations. 

Let us consider the twistor field $g_I(Z)$ and the Lorentz scalars
$(\la^a g_a, \mu^{\dot{a}} g_{\dot{a}}, \pa_a g^a, \pa_{\dot{a}} 
g^{\dot{a}})$ are homogeneous of degree $(0,0,-2,-2)$
since $g_a$ and $g_{\dot{a}}$ are homogeneous of degree $-1$.
By counting of the gauge invariance and the constraint(leading to
the removal of two fields of degree $0$), 
there exist two twistor fields of degree 
$-2$ leading to 
two massless states of helicity 0.
The field $g_{A=1}$
is homogeneous of weight $-1$ describing 
massless states of helicity $\frac{1}{2}$ while
the field $g_{A=2}$
is homogeneous of weight $-3$ describing 
massless states of helicity $-\frac{1}{2}$.
However, this is not allowed in our consideration. 
Then the  
complete structure 
of helicity states 
by the field $g_I(Z)$
are summarized by
\bea
\pa_a g^a(Z) &:& (0, 1 ), 
\;\; (-\frac{1}{2}, 0 ), \;\;
\nonu \\
\pa_{\dot{a}} g^{\dot{a}}(Z) &:& (0, 1), \;\; (-\frac{1}{2}, 0), 
\;\; 
\label{Gtwistor} \\
g_{A=1}(Z) &:& (\frac{1}{2}, 1), \;\; (0, 0), \;\;(-1, 0), 
\;\; (-\frac{3}{2}, -1). 
\nonu 
\eea
Although  in ${\cal N}=4$ case, the massless 
fields described by $g_I(Z)$ possess 
the opposite helicities and conjugate $SU(4)$ representations
from those described by $f^I(Z)$, 
this is not true for our ${\cal N}=1$ case.
We also note that 
the first two entries in each $g_I(Z)$
have the same helicities as the one in ${\cal N}=4$ case \cite{BW}
but the $U(1)_R$ charges are different.
By acting some quantity on $g^I(Z)$ of ${\cal N}=4$
twistor field($I=a$ or
$\dot{a}$) which will change the $U(1)_R$ charges  
of $4$ of state $(0,{\bf 1})$ and $3$ of state $(-\frac{1}{2},
\overline{\bf 4})$ into
$1$ or $0$ respectively, this modified $g_I(Z)$ will give
the above assignments.
Similarly the modified $g_{A=1}(Z)$ that can be obtained 
by acting some quantity on 
$g_{A=1}(Z)$ will change $U(1)_R$ charges
$3$ of state $(\frac{1}{2},\overline{\bf 4})$ and 
$2$ of state $(0, \overline{\bf 10} \oplus {\bf 6})$  
into $1$ or $0$ respectively.  
Since the fermionic coordinates $\theta_{a}^{A}$
has $U(1)_R$ charge 1, it is natural to consider a superspace 
derivative which has $U(1)_R$ charge $-1$, as a multiplicative 
factor.
 
We will verify that the spacetime fields characterized by
the twistor superfields $f^I(Z)$ and $g_I(Z)$  contain
the physical states of ${\cal N}=1$ conformal supergravity 
in section 4.

\section{Linearized spectrum of ${\cal N}=1$ conformal supergravity}
\setcounter{equation}{0}

\indent

The linearized ${\cal N} \leq 4$ conformal supergravity 
\cite{Siegel,BDD,FZ} 
can be described by a chiral field strength superfield
which has $2s$ spinor indices and  
has dimension $s=\frac{1}{2}(4-{\cal N})$.
It contains spins, 
$s, s+\frac{1}{2}, \cdots, s+\frac{\cal N}{2}(=2)$ and
satisfies some constraints.
For example, for ${\cal N}=4$, 
the chiral field strength superfield with no spinor 
indices has the following 
component field expansion \cite{BW}
\bea
{\cal W}^{{\cal N}=4} (x, \theta) &=&
\cdots + (\theta^3)_D^{(abc)} \; (\pa \eta)_{(abc)}^D+
(\theta^3)_{[AB]}^{aC} \; \xi_{aC}^{[AB]} \nonu \\
&& + 
(\theta^4)_{B}^{A(ab)} \; (\pa V)_{(ab)A}^{B} +
(\theta^4)^{(abcd)} \; W_{abcd} + (\theta^4)_{[CD]}^{[AB]} 
\; d^{[CD]}_{[AB]} \nonu \\
&& + (\theta^5)_{C}^{a[AB]} \; \pa_{a\dot{a}} 
\overline{\xi}_{[AB]}^{\dot{a}C} +(\theta^5)^{A(abc)} 
\; (\pa \rho)_{A(abc)}
+
\cdots
\label{n4}
\eea
where the terms with lower $\theta$'s than $\theta^3$ 
and the terms higher $\theta$'s than $\theta^5$
are not written explicitly.
The flat superspace covariant derivatives are given by 
\bea
D^a_A = \frac{\pa}{\pa \theta_{a}^A} + 
\overline{\theta}^{\dot{a}}_A \pa_{a\dot{a}}, \qquad 
\overline{D}_{\dot{a}}^A =\frac{\pa }{\pa 
\overline{\theta}^{\dot{a}}_A}, \qquad a,\dot{a}=1,2 \;\;\;
A=1,2,3,4. 
\nonu
\eea
The chirality of ${\cal W}^{{\cal N}=4}(x,\theta,
\overline{\theta})$ implies 
$\overline{D}_{\dot{a}}^A 
{\cal W}^{{\cal N}=4}(x,\theta,\overline{\theta})=0$ 
and this leads to
an independence of $\overline{\theta}^{\dot{a}}_A$.
In (\ref{n4}), we only considered $\theta_a^A$-expansion:chiral 
Minkowski superspace. 

Let us consider ${\cal N}=1$
chiral field strength superfield characterized by 
three(that is, $2s=4-{\cal N}=3$) spinor indices,
according to above counting, and describe it in terms of 
${\cal N}=4$ chiral superfield ${\cal W}^{{\cal N}=4}(x, \theta)$  
because one can use the results of \cite{BW} directly
once we find out the exact relation between the 
two chiral superfields.
One can construct the following quantity
where the spinor indices $a,b$ and $c$ can be obtained 
from three superspace derivatives with 
contracted $SU(4)$ indices(this is equivalent to take three 
superspace derivatives which are 
irrelevant to ${\cal N}=1$ superspace)
by using the 4th rank epsilon tensor and put the other fermionic
coordinates to zero after differentiation:
\bea
{\cal W}^{{\cal N}=1}_{abc}(x, \theta_1) = 
\epsilon_{1ABC} D^A_a D^B_b D^C_c {\cal W}^{{\cal N}=4}(x,\theta) 
\Big |_{\theta_2=\theta_3=\theta_4=0}
\label{abc}
\eea
which is symmetric in the indices $a,b$ and $c$:
An interchanging of any two superspace derivatives
has minus sign due to the anticommutativity and minus sign 
for the changing of $SU(4)$ indices with an antisymmetric 
epsilon tensor.

Let us compute the 
right hand side of (\ref{abc})
by acting the superspace derivatives on 
(\ref{n4}) and putting $\theta_i (i=2,3,4)$ to zero.
One can easily see that the terms which are not written 
explicitly in (\ref{n4})
do not contribute in this procedure.
When three $D$'s which has the $SU(4)$ indices $2,3$ and $4$
are acting on the object which contains more than
six $\theta$'s, we are left with an object with more 
than three $\theta$'s 
and those $\theta$'s have $\theta_i$ where $i\neq 1$. By putting 
$\theta_i (i=2,3,4)$ to zero, the results do not have any
contributions. On the other hand, when three $D$'s are acting on 
the object which has two $\theta$'s at most, there exists 
a single $D$
we have not used. Therefore, there are no contributions.
Note that the second term of $D^a_A$ above does not contribute at all
when acting on some function appearing in the 
${\cal W}^{{\cal N}=4}(x,\theta)$
because it contains $\overline{\theta}^{\dot{a}}_A $ and at the final
expression we put this to zero also.

Now let us first consider 
the cubic term in $\theta$ transforming $\bf 4$ 
of $SU(4)$ with $(\pa \eta)_{(abc)}^D$. 
One can write explicitly as follows
\bea
\epsilon_{1ABC} D^A_a D^B_b D^C_c  \; (\theta^3)^{D(efg)}
=\epsilon_{1ABC} 
 \epsilon^{DEFG}
D^A_a D^B_b D^C_c  \; \theta ^e_E \theta^f_F
\theta^g_G.
\label{CONT1}
\eea
Then this will lead to 
the product of delta functions 
$\delta^D_1 \delta^e_a \delta^f_b \delta^g_c$ 
and provides us with  
$(\pa \eta)_{(abc)}^{D=1}$.
On the other hand, when three superspace derivatives act on  
$(\pa \eta)_{(abc)}^D$, 
since the second term of a superspace derivative $D^a_A$ does not
contribute 
at all, there is no contribution.

Similarly the next term in cubic $\theta$ 
can be written as, by applying the irreducible
$D$-operators in ${\cal N}=4$ superspace \cite{gs,superspace} to 
the fermionic coordinates,
\bea
\epsilon_{1ABC} D^A_a D^B_b D^C_c  \; (\theta^3)_{[DE]F}^{d}
=\epsilon_{1ABC}  \epsilon_{ef} D^A_a D^B_b D^C_c  \; 
\theta^d_{[D} \theta^e_{E]} \theta^f_F 
\nonu
\eea
where the 2nd rank antisymmetric tensor 
$\ep_{ab}$ is defined in \cite{WB,Witten}. 
The spinor indices can be raised or lowered by using this and its
inverse $\ep^{ab}$.
By calculating the above quantity  explicitly, 
it turns out there are six terms 
which vanish from the antisymmetric property of 4th rank and 
2nd rank epsilon tensors. 

Let us move on the quartic term in $\theta$ transforming 
$\bf 15$ of $SU(4)$:
\bea
\epsilon_{1ABC} D^A_a D^B_b D^C_c  \; 
(\theta^4)_{D}^{E(de)} = \epsilon_{1ABC} 
\epsilon^{EFGH} \epsilon_{fg}
D^A_a D^B_b D^C_c  \; 
\theta^f_D \theta^g_F \theta^d_G \theta^e_H.
\label{quartic}
\eea
The contraction between 4th rank epsilon tensors
gives rise to the product of two delta functions after acting 
three superspace derivatives on the $\theta$'s.
When the index $D$ and index $E$ are equal to each other $D=E$, 
there exists a term like as 
$\ep_{cd} \theta^d_1 (\pa V)^A_{(ab)A}$. However, 
since $V_{\mu A}^A=0$ \cite{BW} that is nothing but irreducibility
condition
of $SU(4)$ tensor, 
all these contributions become zero.
When they are different from each other $D\neq E$, 
the $SU(4)$ indices of a single remaining $\theta$
can be either 2,3 or 4. Therefore, these contributions are zero after
we set $\theta_2=\theta_3=\theta_4=0$ eventually.

By writing the next term transforming a singlet 
${\bf 1}$ of $SU(4)$ as 
\bea
\epsilon_{1ABC} D^A_a D^B_b D^C_c  \; (\theta^4)^{(defg)}
=\epsilon_{1ABC} 
\epsilon^{DEFG}
D^A_a D^B_b D^C_c  
\; \theta^d_D \theta^e_E \theta^f_F \theta^g_G,
\label{CONT2}
\eea
then the $\theta$ that are left with contains $SU(4)$ index $1$ and 
the nontrivial result is given by 
$\theta^d_1 W_{(abc)d}$.

Moreover, the last term in quartic $\theta$ transforming 
${\bf 20}^{\prime}$ 
of $SU(4)$ can be written as
\bea
\epsilon_{1ABC} D^A_a D^B_b D^C_c  \; (\theta^4)_{[FG]}^{[DE]} =
 \epsilon_{1ABC} 
 \epsilon^{DEHI}
\epsilon_{de} \epsilon_{fg}
D^A_a D^B_b D^C_c  \;  \theta_H^d 
\theta_F^e \theta_I^f \theta_G^g.
\nonu
\eea
In this case, twenty four terms are exactly cancelled each other 
by the antisymmetric property of epsilon tensors.
 
Now we turn to next order term in $\theta$ transforming $\bf 20$ 
of $SU(4)$ and decompose as follows:
\bea
\epsilon_{1ABC} D^A_a D^B_b D^C_c \;  (\theta^5)_{Fd}^{[DE]}
=\epsilon_{1ABC} 
\epsilon^{DEGH}
D^A_a D^B_b D^C_c 
\left[(\theta^4)_{FGde} \theta^{e}_H \right].
\nonu
\eea
From the previous consideration (\ref{quartic}), 
there is no contribution 
when the three
superspace derivatives are acting on only the first factor 
$(\theta^4)_{FGde}$. Then the nontrivial contributions should   
contain a superspace derivative acting on the second factor 
$\theta^{e}_H$. Of course, the remaining two superderivatives
act on the first factor.
Then the results can be written as 
the contractions between three 4th rank epsilon tensors
together with the quadratic in $\theta$'s.
It is easy to see that one of the 
$SU(4)$ indices in $\theta$'s
can be either 2,3 or 4.
Therefore, there are no nonzero contributions after we put 
$\theta_i=0$ where $i=2,3,4$. 

Let us consider the next term in fifth order in $\theta$
transforming $\bf 4 $ of $SU(4)$
with $(\pa \rho)_{A(abc)}$ and write explicitly as follows
\bea
\epsilon_{1ABC} D^A_a D^B_b D^C_c \; (\theta^5)^{D(def)} =
\epsilon_{1ABC} D^A_a D^B_b D^C_c \left[
(\theta^4)^{DE(de)} \theta_E^f\right].
\label{CONT3}
\eea
One can do this similarly.
When the $SU(4)$ index 
$D$ is equal to $A,B$ or $C$ among 36 terms, the index $D$ can be 
$2,3$ or 4 and one of 
the $SU(4)$ indices in the remaining two $\theta$'s
is equal to the index $D$. 
This implies that 18 terms do not contribute.
However, when the index $D$ appears in the
remaining two $\theta$'s, then 
this will provide the following nonzero contribution 
$\ep_{de} \theta^d_1 \theta^e_1 (\pa \rho)_{1(abc)}$.  

After we combine all the nonzero contributions (\ref{CONT1}), 
(\ref{CONT2}) and (\ref{CONT3})
together,
the relation (\ref{abc}) reads 
\bea
{\cal W}_{abc}^{{\cal N}=1}(x^{d\dot{d}},\theta_d^1) = 
(\pa \eta)_{(abc)} +  \theta^{d}_1 \left[ C_{(abc)d} + 
\ep_{ab} (\pa A)_{cd} + \ep_{ac} (\pa A)_{db}+
\ep_{ad} (\pa A)_{bc} 
\right] + \theta^d_1 \theta_{d1} 
(\pa \rho)_{(abc)}
\nonu
\eea 
where the $W_{(abc)d}$ is decomposed as 
$C_{abcd}$ which is symmetric in all indices plus $A_{cd}$
which is symmetric in indices \footnote{
Here we use simplified notation as \cite{BW} and denote 
$(\pa A)_{(ab)}  =  \pa_{\mu} A_{\nu} (\sigma^{\mu\nu})_{(ab)}$, 
$(\pa \eta)_{(abc)}  =  
\pa_{\mu} \eta_{\nu(a} (\sigma^{\mu\nu})_{bc)}$,
$(\pa \rho)_{(abc)}  = 
\pa_{\mu} \rho_{\nu(a} (\sigma^{\mu\nu})_{bc)}$,
and $\rho_{\nu a}  =  \sigma^{\mu}_{a\dot{a}} \left( 
\pa_{\mu} \overline{\eta}_{\nu }^{\dot{a}} -\pa_{\nu}
\overline{\eta}_{\mu }^{\dot{a}} + \frac{1}{2} \ep_{\mu \nu \tau 
\kappa} \pa^{\tau} \overline{\eta}^{\kappa \dot{a}}
\right)$.}.
This is exactly the same as a superfield approach \cite{FZ} to
${\cal N}=1$ conformal supergravity where the complete
analysis of the linearized ${\cal N}=1$ conformal supergravity was 
given.
What we have done so far is that for given ${\cal N}=4$
superfield (\ref{n4}), we computed the right hand side of (\ref{abc})
and the right field contents of ${\cal N}=1$ conformal
supergravity were selected without imposing on the vanishing of
extra field contents of ${\cal N}=4$ theory by hand, by doing the 
superspace derivative algebra on the $SU(4)$ 
tensorial structure of the 
fermionic coordinates.  

The action for ${\cal N}=1$ conformal supergravity 
at the linearized level \cite{FZ,FT} can be written as
\bea
S & = & \int d^4 x \left(
\int d^8 \theta \; {\cal W}_{{\cal N}=4}^2 \right) 
\Big |_{\theta_i=0} 
 \rightarrow   
\int d^4 x \int d^2 \theta_1 \; 
\left[{\cal W}_{abc}^{{\cal N}=1}(x,\theta_1) \right]^2.
\nonu
\eea
By computing the $\theta_1$-integrals with the explicit 
form for ${\cal W}_{abc}^{{\cal N}=1}(x^{d\dot{d}},
\theta_d^1)$ above, 
this action
leads to the sum of following component Lagrangian 
\bea
{\cal L}_1=  (\pa A)_{(ab)} (\pa A)^{(ab)}, \qquad
{\cal L}_{\frac{3}{2}} = (\pa \eta)_{(abc)} 
(\pa \rho)^{(abc)}, \qquad
{\cal L}_2= C_{abcd} C^{abcd}.
\nonu
\eea 
The quartic and higher order fermionic terms 
can be found in \cite{KTv}.
See also recent review paper on the conformal supergravity 
\cite{vanN}.
The linearized ${\cal N}=1$ conformal supergravity can be described
off-shell
by a chiral field strength superfield  
${\cal W}_{abc}^{{\cal N}=1}(x^{d\dot{d}},\theta_d,
\overline{\theta}^{\dot{d}})$ 
that satisfies the constraint
$
\pa_{\dot{a}}^{\;\;a} D^b  {\cal W}^{{\cal N}=1}_{abc} = 
\pa_{a}^{\;\; \dot{a}} \overline{D}^{\dot{b}}  
\overline{{\cal W}}^{{\cal N}=1}_{\dot{a} \dot{b} \dot{c}}
$  \cite{FZ,FT}.
Then the condition for 
${\cal W}_{abc}^{{\cal N}=1}(x,\theta,\overline{\theta})$ 
to be chiral, $\overline{D}^{\dot{d}} 
{\cal W}^{{\cal N}=1}_{abc}(x^{e\dot{e}},\theta_e,
\overline{\theta}^{\dot{e}})=0$
implies that ${\cal W}^{{\cal N}=1}_{abc}(x^{e\dot{e}},\theta_e,
\overline{\theta}^{\dot{e}})$ does not depend on 
$\overline{\theta}^{\dot{e}}$.

To determine the helicities for these fields, 
it is  necessary to solve the higher derivative equations of motion:
the spin $\frac{3}{2}$-field has 
third-order derivatives and the spin 
$2$-field has fourth-order derivatives.

For  two gravitinos $\eta_{\mu }^{a}$, 
the linearized equation of motion \footnote{
In the original paper by \cite{FZ}, the equation of 
motion for the Rarita-Schwinger field in the four component
spinor notation reads 
$\left(\gamma \cdot \pa \Box \delta_{\mu \nu} - 
\gamma \cdot \pa \pa_{\mu}
\pa_{\nu} -\frac{1}{2} \ep_{\nu \mu \rho \sigma} \gamma_5
\gamma^{\sigma}
\pa^{\rho} \Box \right) \eta^{\nu}=0$ and one can convert this 
into the above two component spinor notation \cite{WB}.} is
given by \cite{BW} 
\bea
\pa^{e\dot{b}} 
\pa^{a (\dot{c}} \pa^{\dot{a}) b} 
\sigma^{\mu}_{b\dot{b}} \eta_{\mu a} =0 
\nonu
\eea
and this implies 
\bea
\eta_{\mu a} = \sigma_{\mu}^{b\dot{b}} \int 
 d^4 k \de(k^2) e^{ik\cdot x} 
\left[ \pi_a \pi_b \tau_{\dot{b}} \left(
\eta_{-\frac{3}{2}}(k) + i
\frac{x_0}{k_0}\eta_{-\frac{3}{2}}^{\prime}(k) \right)
+\pi_{(a} \widetilde{\tau}_{b)} \tau_{\dot{b}} 
\eta_{-\frac{1}{2}}(k) +
\widetilde{\tau}_a \widetilde{\tau}_b 
\widetilde{\pi}_{\dot{b}} \eta_{\frac{3}{2}}(k)  \right].
\label{etacsg}
\eea
Here we used a pair of spinors $\pi, \widetilde{\pi}$ such that
$k^{a\dot{a}}=\pi^a \widetilde{\pi}^{\dot{a}}$
and a pair of spinors $\tau, \widetilde{\tau}$
such that $\pi^a \widetilde{\tau}_a =1$ and 
$\widetilde{\pi}^{\dot{a}} 
\tau_{\dot{a}}=1$. Both 
$\eta_{-\frac{3}{2}}(k)$  and $\eta_{-\frac{3}{2}}^{\prime}(k)$ 
which 
are independent fields combine
to a `dipole' field of helicity $-\frac{3}{2}$ and spacetime
translations act in undiagonalizable reflecting the nonunitarity of
the theory.

For $A_{\nu}$,
the equation of motion from the action is given by 
\footnote{
One can also write this in different form
$\left( \delta^{\nu \sigma} \Box -\pa^{\nu} \pa^{\sigma} \right) 
A_{\sigma}
=0$ from the property of $\sigma^{\mu \nu}$ matrix \cite{WB}.}
\bea
\pa_{\mu} \pa_{\rho} 
(\sigma^{\mu \nu})_{(ab)} 
(\sigma^{\rho \sigma})^{(ab)} 
 A_{\sigma}=0
\nonu
\eea
and one can obtain 
\bea
A_{\mu} =
\sigma_{\mu}^{a\dot{a}} 
\int d^4 k \de(k^2) e^{ik\cdot x} \left[ \pi_a \tau_{\dot{a}} 
 A_{-1 }(k) +
\widetilde{\tau}_a \widetilde{\pi}_{\dot{a}}  
A_{1 }(k)  \right].
\label{acsg}
\eea
For graviton $e_{\mu a \dot{a}}$, the equation of motion with 
fourth order derivative is 
given by \cite{BW}
\bea
\pa^{\dot{a} (c} \pa^{d) \dot{b}} \pa^{a(\dot{c}} \pa^{\dot{d})b} 
\sigma_{b\dot{b}}^{\mu} e_{\mu a \dot{a}} =0
\nonu
\eea
and the solution 
\footnote{The conformal invariant equation for 
spin $2$-field can be
written as $\frac{1}{2} P_{2 \rho \sigma}^{\mu \la} 
\Box^2 h_{\rho \sigma}$ with symmetric metric tensor 
$h_{\mu \nu}$ where $P_{2 \rho \sigma}^{\mu \nu}$ is a spin 2 
projector \cite{FT}:
$P_{2 \rho \sigma}^{\mu \nu}=\Pi_{(\rho}^{\mu} \Pi_{\sigma)}^{\nu} -
\frac{1}{3} \Pi_{\rho \sigma} \Pi^{\mu \nu} $ and
$\Pi_{\nu}^{\mu}=\de_{\nu}^{\mu} -
\pa^{\mu}
\Box^{-1} \pa_{\nu}$.} 
is 
\bea
e_{\mu a \dot{a}} & = & \sigma_{\mu}^{b\dot{b}} \int
d^4 k \de(k^2) e^{ik\cdot x} \left[\pi_a \pi_b \tau_{\dot{a}}
  \tau_{\dot{b}}
\left( e_{-2}(k) +i \frac{x_0}{k_0} e_{-2}^{\prime}(k) \right)+
\pi_{(a} \widetilde{\tau}_{b)} \tau_{\dot{a}} \tau_{\dot{b}} 
e_{-1}(k)
\right. \nonu \\
&& \left. + 
\widetilde{\tau}_a \widetilde{\tau}_b \widetilde{\pi}_{(\dot{a}}
\widetilde{\tau}_{\dot{b})}e_1(k) 
 + \widetilde{\tau}_a \widetilde{\tau}_b \widetilde{\pi}_{\dot{a}}
\widetilde{\pi}_{\dot{b}} \left( e_2(k) +i \frac{x_0}{k_0}
  e_{2}^{\prime}(k) 
\right) \right].
\label{spin2csg}
\eea
In this case, there exist two dipole fields of helicity $-2$ and $2$.
The linearized Weyl tensor $C_{\mu\nu\rho\sigma}$ 
can be defined through a Riemann
tensor $R_{\mu\nu\rho\sigma}$ written in terms of 
second derivative of symmetric metric tensor, 
its contracted expressions 
$R_{\nu\sigma}, R_{\mu\sigma},
R_{\nu\rho},
R_{\mu\rho}$ and $R_{\mu}^{\mu}$ \cite{FZ}  
and $C_{abcd}$ is the spinorial equivalent of this Weyl conformal 
tensor $C_{\mu\nu\rho\sigma}$: 
it is called the gravitational spinor in \cite{Penann}.

\begin{table}
\begin{center} 
\begin{tabular}{cccc}  
  \hline
\vspace{0.2cm}   
States in ${\cal N}=1$ CSG & $U(1)_R$ charge & Helicity  
& Twistor superfields \\ 
\hline
\vspace{0.3cm}
$\eta_{\mu }^{a}$   & $1$ &     $-\frac{3}{2}^{\prime} $     &
$\epsilon_{1ABC} D^A_a D^B_b D^C_c  \; g_{\dot{a}}(Z) 
\Big |_{\theta_2=\theta_3=\theta_4=0}$ 
\\ 
\vspace{0.3cm}
&$1$  & $-\frac{3}{2} $ &
$\epsilon_{1ABC} D^A_a D^B_b D^C_c  \; g_{d}(Z) 
\Big |_{\theta_2=\theta_3=\theta_4=0}$
\\
\vspace{0.3cm}
& $1$ & $-\frac{1}{2}$ & 
$\epsilon_{12BC} D^B_b D^C_c  \; g_{D=1}(Z) 
\Big |_{\theta_3=\theta_4=0}$ 
\\ 
\vspace{0.3cm}
&$1$ & $\frac{3}{2}$  &
$ f^{A=1}(Z) \Big |_{\theta_3=\theta_4=0}$ 
\\
\hline
\vspace{0.3cm}
$A_{\mu}$   & $0$ &     $1$     
&$ f^{A=1}(Z) \Big |_{\theta_3=\theta_4=0}$    
\\
\vspace{0.3cm}
&$0$ &$-1$ & $\epsilon_{12BC} D^B_b D^C_c  \; 
g_{D=1}(Z) 
\Big |_{\theta_3=\theta_4=0}$ 
\\
\hline
\vspace{0.3cm}
$e^{a \dot{a}}_{\mu}$   & $0$ &    $2$  &
$f_{\dot{a}}(Z) 
\Big |_{\theta_2=\theta_3=\theta_4=0}
$                 
\\
\vspace{0.3cm}
&$0$ &$2^{\prime}$ & $\la^a f_a(Z) 
\Big |_{\theta_2=\theta_3=\theta_4=0}  $ 
\\ 
\vspace{0.3cm}
&$0$ & $1$ & 
$f^{A=1}(Z) \Big |_{\theta_3=\theta_4=0} $
\\
\vspace{0.3cm}
&$0$ &$-1$ & 
$\epsilon_{12BC} D^B_b D^C_c  \; g_{D=1}(Z) 
\Big |_{\theta_3=\theta_4=0}$ 
\\
\vspace{0.3cm}
&$0$ &$-2$ &
$\epsilon_{1ABC} D^A_a D^B_b D^C_c  \; g_{d}(Z) 
\Big |_{\theta_2=\theta_3=\theta_4=0}$
\\
\vspace{0.3cm}
&$0$ &$-2^{\prime}$ &
$\epsilon_{1ABC} D^A_a D^B_b D^C_c  \; g_{\dot{a}}(Z) 
\Big |_{\theta_2=\theta_3=\theta_4=0}$ 
\\
\hline
\vspace{0.3cm}
$\overline{\eta}^{\dot{a}}_{\mu }$   & $-1$ &
$\frac{3}{2}$    &
$f_{\dot{a}}(Z) 
\Big |_{\theta_2=\theta_3=\theta_4=0} $
\\
\vspace{0.3cm}
&$-1$ & $\frac{3}{2}^{\prime}$   
& $\la^a f_a(Z) 
\Big |_{\theta_2=\theta_3=\theta_4=0} $
\\ 
\vspace{0.3cm}
&$-1$ & $\frac{1}{2}$ & 
$f^{A=1}(Z) \Big |_{\theta_3=\theta_4=0} $
\\ 
\vspace{0.3cm}
&$-1$ & $-\frac{3}{2}$ & 
$\epsilon_{12BC} D^B_b D^C_c  \; g_{D=1}(Z) 
\Big |_{\theta_3=\theta_4=0}$ 
\\
\hline
\end{tabular} 
\begin{center}
\caption{\sl The $U(1)_R$ charges, helicities of 
physical states in ${\cal N}=1$ conformal 
supergravity(CSG) 
in four dimensions. In the last column the relevant
twistor
superfields are given through 
(\ref{twistoradot}), 
(\ref{twistorA}), (\ref{twistora}) and (\ref{twistorf}) where
$f^I(Z), g_I(Z)$ and $D_a^A$ are ${\cal N}=4$ objects. There are
four dipole fields (two states with the same helicity) 
of helicities 
$-2, -\frac{3}{2}, \frac{3}{2},2$ in which the spacetime translations
act in nondiagonalizable. The dependence on the fermionic coordinate 
$\chi$ of weight $3$ of ${\bf WCP}^{3|2}$ appears only in both 
two superspace derivatives acting on $g_{A=1}(Z)$ and 
$f^{A=1}(Z)$ at $\theta_3=\theta_4=0$. }
\end{center}
\end{center}
\end{table}

We will prove that 
the physical states of ${\cal N}=1$ conformal 
supergravity in four dimensions 
can be obtained from the spacetime fields described by the twistor 
fields  $f^I(Z)$ and $g_I(Z)$.
Let us start with the identification of 
the chiral superfield ${\cal W}_{abc}^{{\cal N}=1}(x,\theta_1)$
in the linearized conformal supergravity with
the twistor fields of ${\cal N}=1$ as follows:
\bea
{\cal W}_{abc}^{{\cal N}=1}(x,\theta_1) = 
\left(\epsilon_{1ABC} D^A_a D^B_b D^C_c 
\int _{{\bf D}_{x,\theta}} 
g_I dZ^I \right) \Big |_{\theta_2=\theta_3=\theta_4=0}, 
\nonu
\eea
where the $Z^I$ are functions of $\la^a$ for fixed $x$ and 
$\theta$ through the twistor equations \cite{Witten}
\bea
dZ^I =(d \la^a, d \mu^{\dot{a}}, 
d\psi^A) =(d \la^a, d \la_b x^{b\dot{a}}, d \la_c \theta^{cA})
\nonu
\eea
and ${\bf D}_{x,\theta}$ is the curve with moduli $x$ and $\theta$.
We will perform the action of superspace derivatives on the twistor 
field $g_I(Z)$. 
Here we used the twistor equations
\bea
(\mu^{\dot{a}}, \psi^A)=(x^{a\dot{a}} \la_a, \theta^{Aa} \la_a).
\nonu
\eea

Let us write the twistor fields  $g_I(Z)$ in terms of 
${\cal W}_{abc}^{{\cal N}=1}(x,\theta_1)$ that has its physical 
fields, $A_{\mu}, \eta_{\mu}^a$ and $e_{\mu}^{ a \dot{a}}$.
Our strategy here is to take ${\cal N}=4$ twistor fields
and use the relation between our ${\cal N}=1$ chiral superfield
and ${\cal N}=4$ superfield (\ref{abc}).
Then the ${\cal N}=1$ twistor fields can be read off from the 
${\cal N}=4$ twistor fields as we will see. 
As observed in section 3, we will see that the super derivative
plays a role of changing $U(1)_R$ charge. 
From the ${\cal N}=4$ twistor field results \cite{BW}
\bea
g_{\dot{a}}^{{\cal N}=4}(Z) 
= \cdots   -i\frac{\la^a \si_{a\dot{a}}^0}{k^0} 
\left[ (\psi^3)_A \hat{\eta}^
{\prime A}_{-\frac{3}{2}} + 
(\psi^4) \hat{e}^{\prime}_{-2} \right]
\nonu
\eea
where the terms with lower $\psi$'s than $\psi^3$ are not written 
explicitly,
the relation turns out \footnote{
One can compute this by using 
$ (\psi^3)^D  = \ep^{DEFG} \theta^e_E \theta^f_F \theta^g_G 
\la_e \la_f \la_g $
and 
$\psi^4 = \ep^{DEFG} \theta^d_D \theta^e_E \theta^f_F 
\theta^g_G \la_d \la_e \la_f \la_g$.}
\bea
\epsilon_{1ABC} D^A_a D^B_b D^C_c  \; g_{\dot{a}}^{{\cal N}=4}(Z) 
\Big |_{\theta_2=\theta_3=\theta_4=0}
= -i\frac{\la^d \si_{d\dot{a}}^0}{k^0}
\la_a \la_b \la_c \left[\hat{\eta}^
{\prime }_{-\frac{3}{2}} + \psi \hat{e}^{\prime}_{-2}   
\right].
\label{twistoradot}
\eea
Here 
$\hat{\eta}^
{\prime }_{-\frac{3}{2}}$ corresponding to 
$\hat{\eta}^
{\prime A=1 }_{-\frac{3}{2}}$ of ${\cal N}=4$
twistor field 
represents the twistor field of 
$GL(1)$ charge  $-5$ for the spacetime field $\eta^
{\prime }_{-\frac{3}{2}}(k)$ (\ref{etacsg}) 
of helicity $-\frac{3}{2}$ and
$ \hat{e}^{\prime}_{-2} $ 
corresponding to 
$ \hat{e}^{\prime}_{-2} $ 
 of ${\cal N}=4$
twistor field 
represents the twistor field of
$GL(1)$ charge $-6$ for the spacetime field 
$e^{\prime}_{-2}(k)$ 
(\ref{spin2csg}) of helicity $-2$.
That is, for example, 
there exists a relation 
\bea
\int d^4 k \de(k^2) e^{ik\cdot x}
{\eta}^
{\prime }_{-\frac{3}{2}}(k) = \int d \la^a \la_a \hat{\eta}^
{\prime }_{-\frac{3}{2}} = {\eta}^
{\prime }_{-\frac{3}{2}}(x).
\nonu
\eea
We also used the fact that there exists a twistor equation 
$\psi^{A=1}=\psi = \theta^{A=1,a} \la_a$.
In section 3, the helicity and $U(1)_R$ charge by the state 
$\pa_{\dot{a}} 
g^{\dot{a}}(Z)$ are given by $(0,1)$ and $(-\frac{1}{2},0)$.
The product of $\la_a \la_b \la_c \hat{\eta}^
{\prime }_{-\frac{3}{2}}$ in (\ref{twistoradot}) corresponds to 
$(0,1)$ twistor field while
 $\la_a \la_b \la_c \hat{e}^{\prime}_{-2}$ corresponds to
$(-\frac{1}{2}, 0)$ twistor field. 
The helicities are changed by the presence of 
 three $\la$'s
(each of them has a helicity $\frac{1}{2}$). Therefore, 
the state $(-\frac{3}{2}, 1)$ which shows an opposite helicity and 
an opposite $U(1)_R$ charge 
from those described by $\mu^{\dot{a}} f_{
\dot{a}}(Z)$  corresponds to $\hat{\eta}^
{\prime }_{-\frac{3}{2}}$ and the state $(-2,0)$ corresponds to
$\hat{e}^{\prime}_{-2}$. 

Next, from the relation \cite{BW}
\bea
g_{A}^{{\cal N}=4}(Z) 
= \cdots + (\psi^2)_{[AB]} \hat{\eta}^B_{-\frac{1}{2}}+
(\psi^2)_{[BC]} \hat{\xi}^{[BC]}_{-\frac{1}{2} A} 
+ (\psi^3)_A \hat{e}_{-1} + (\psi^3)_B 
\hat{V}^B_{-1 A} + (\psi^4) \hat{\overline{\eta}}_{-\frac{3}{2} A}
\nonu
\eea
which was obtained by acting $\frac{\pa}{\pa \theta_a^A}$ on ${\cal
  W}^{{\cal N}=4}(x,\theta)$, 
one gets, by using two superspace derivatives rather than three 
because one superspace derivative is used at the level of 
${\cal N}=4$ theory already, 
\bea
\epsilon_{12BC} D^B_b D^C_c  \; g_{D=1}^{{\cal N}=4}(Z) 
\Big |_{\theta_3=\theta_4=0}
= 
\la_b \la_c \left[\hat{\eta}_{-\frac{1}{2}} + 
\psi \hat{A}_{-1} + \chi \hat{e}_{-2} + 
\psi \chi \hat{\overline{\eta}}_{-\frac{5}{2}}   
\right].
\label{twistorA}
\eea
The $\chi$ can be written as $\la^a \la^b \la^c \theta^2_{abc}$ 
in general \footnote{One can express $\psi$ and $\chi$ in terms of 
$\theta_A^a$. One factorizes ${\bf CP}^{3|4}$ with respect to 
two other fermionic vector fields $N_1= \la_2 
\frac{\pa}{\pa \psi^2} -\la_1 \frac{\pa}{\pa \psi^3}$ and $
N_2= \la_1 
\frac{\pa}{\pa \psi^4} -\la_2 \frac{\pa}{\pa \psi^3}
$ where $\psi^A =\theta^{Aa} \la_a$. 
These vector fields allow us to reduce $\Pi{\cal O}(1) \oplus 
\Pi{\cal O}(1) \oplus \Pi{\cal O}(1)$ to $\Pi{\cal O}(3)$.
Remember that the fermionic coordinates $\psi$ and $\chi$
take values in the holomorphic line bundles ${\cal O}(1)$ and 
${\cal O}(3)$ respectively \cite{PW,AHN}. 
Here $\Pi$ is a parity changing operator which changes  
the parity of the fiber coordinates when acting on a fiber 
bundle \cite{PS}.  
Then the fermionic coordinates invariant with respect to 
$N_1$ and $N_2$(i.e., they are annihilated by $N_1$ and $N_2$) 
of 
${\bf WCP }^{3|2}$
are given by $\psi=\psi^1$ and $\chi=\la_1 \la_1 \psi^2+
\la_1 \la_2 \psi^3 + \la_2 \la_2 \psi^4 \equiv \la_a \la_b \la_c 
\theta_2^{abc}$. We thank A.D. Popov for pointing out this.
\label{chifoot}} and can be written as a linear combination of
$\theta_i(i=2,3,4)$ with the coefficient that is cubic in $\la$'s.
In the above computation, we put $\theta_3=\theta_4=0$ at the 
end.

Let us describe how each term in (\ref{twistorA}) comes from 
those in ${\cal N}=4$ contents.
The twistor field $\hat{\eta}_{-\frac{1}{2}}$ corresponding to 
$ \hat{\xi}^{[B=3,C=4]}_{-\frac{1}{2}, A=1}$ of ${\cal N}=4$
twistor field
describes the twistor field of $GL(1)$ charge $-3$
for the spacetime field $\eta_{-\frac{1}{2}}(k)$ of helicity 
$-\frac{1}{2}$. 
The contribution from 
$ (\psi^2)_{[AB]}  \hat{\eta}^B_{-\frac{1}{2}}$ 
when the $SU(4)$ index $A$ is equal to $1$
vanishes because the two superspace derivatives in 
(\ref{twistorA}) 
pick up only the $SU(4)$ indices 
$3$ and $4$. 
The twistor field
$\hat{A}_{-1}$ corresponding to 
$\hat{V}^{B=2}_{-1, A=1}$ of ${\cal N}=4$ twistor field
denotes the twistor field of $GL(1)$ charge $-4$
for the spacetime field $A_{-1}(k)$ in (\ref{acsg}) 
of helicity $-1$. The twistor field
$\hat{e}_{-1}$ where $\hat{e}_{-2}$ multiplied by 
two $\la$'s 
is equal to $\hat{e}_{-1}$ of ${\cal N}=4$ 
twistor field(Note that $\chi=\la_1 \la_1 \psi^2$ at 
$\theta_3=\theta_4=0$)
describes 
  the twistor field of $GL(1)$ charge $-6$
for the spacetime field $e_{-1}(k)$ of helicity 
$-1$. Here we used the same notation $\hat{e}_{-1}$ 
as ${\cal N}=4$ field contents.
  Finally the twistor field 
$\hat{\overline{\eta}}_{-\frac{3}{2}}$ where 
the quantity $\hat{\overline{\eta}}_{-\frac{5}{2}}$ multiplied 
by two $\la$'s corresponds to 
 $\hat{\overline{\eta}}_{-\frac{3}{2}, A=1}$
of ${\cal N}=4$ twistor field  denotes 
the twistor field of $GL(1)$ charge $-7$
for the spacetime field $\overline{\eta}_{-\frac{3}{2}}(k)$ 
of helicity 
$-\frac{3}{2}$.    
The expression $\la_b \la_c \hat{\eta}_{-\frac{1}{2}}$
denotes the state $(\frac{1}{2},1)$ of section 3.
Similarly $\la_b \la_c \hat{A}_{-1}$ corresponds to 
$(0,0)$ twistor field, $\la_b \la_c \hat{e}_{-2}$ does $(-1,0)$
twistor
field and   $\la_b \la_c 
\hat{\overline{\eta}}_{-\frac{5}{2}}$ describes 
$(-\frac{3}{2},-1)$
twistor field.   
All these twistor fields have their counterparts 
in $f^{A=1}(Z)$ below that have opposite helicities and 
$U(1)_R$ charges.

From the expression
\bea
g_a^{{\cal N}=4}(Z) = \cdots + \la_a \left[ (\psi^3)_A 
\hat{\eta}^A_{-\frac{3}{2}} + (\psi^4) \hat{e}_{-2} \right],
\nonu
\eea
one can compute 
\bea
\epsilon_{1ABC} D^A_a D^B_b D^C_c  \; g_{d}^{{\cal N}=4}(Z) 
\Big |_{\theta_2=\theta_3=\theta_4=0}
=\la_d \la_a \la_b \la_c \left[ \hat{\eta}_{-\frac{3}{2}} 
+ \psi \hat{e}_{-2}   \right].
\label{twistora}
\eea
One can see that $\hat{\eta}_{-\frac{3}{2}}$ 
 corresponding to 
$\hat{\eta}^
{A=1 }_{-\frac{3}{2}}$ of ${\cal N}=4$
twistor field 
denotes the twistor field of 
$GL(1)$ charge  $-5$ for the spacetime field 
$\eta_{-\frac{3}{2}}(k)$ (\ref{etacsg}) 
of helicity $-\frac{3}{2}$ and
$ \hat{e}_{-2} $ 
corresponding to
$ \hat{e}_{-2} $ 
 of ${\cal N}=4$
twistor field 
denotes the twistor field of
$GL(1)$ charge $-6$ for the spacetime field 
$e_{-2}(k)$ 
(\ref{spin2csg}) of helicity $-2$.
Similarly, 
the state $(-\frac{3}{2}, 1)$ of section 3
corresponds to $\hat{\eta}_{-\frac{3}{2}}$(or 
$\la_a \la_b \la_c \hat{\eta}_{-\frac{3}{2}}$ has a helicity and 
$U(1)_R$ charge by 
$(0,1)$) and 
the state $(-2,0)$  corresponds to
$\hat{e}_{-2}$. As observed above, one can see here the 
counterpart of these states in $\la^a f_a(Z)$ below.

One obtains the $f^I(Z)$ from the dual field 
$\widetilde{g}^I(\overline{Z})$ from the ${\cal N}=4$
description \footnote{For convenience, we list here 
the ${\cal N}=4$ description by \cite{BW} up to a quadratic 
in $\psi^A$:
$
\la^a f_a^{{\cal N}=4}(Z) 
 =  \hat{e}_2^{\prime} +\psi^A 
\hat{\overline{\eta}}_{\frac{3}{2} A}^{\prime} +
(\psi^2)_{[AB]} \hat{\overline{T}}^{\prime [AB]}_1 +
\cdots$, 
$f^{A}_{{\cal N}=4}(Z) 
 =  \hat{\eta}_{\frac{3}{2}}^A + 
\psi^B \hat{V}^B_{1A} + \psi^A \hat{e}_1  + (\psi^2)^{[BC]}
\hat{\overline{\xi}}^A_{\frac{1}{2}[BC]}  
+ (\psi^2)^{[AB]} \hat{\overline{\eta}}_{\frac{1}{2}B} + \cdots$, 
and $
f_{\dot{a}}^{{\cal N}=4}(Z)  
 = \pa_{\dot{a}} ( \hat{e}_2 + \psi^A 
\hat{\overline{\eta}}_{\frac{3}{2}A} + (\psi^2)_{[AB]} 
\hat{\overline{T}}^{[AB]}_1 ) + \cdots$.  }:
\bea
\la^a f_a^{{\cal N}=4}(Z) 
\Big |_{\theta_2=\theta_3=\theta_4=0} 
& = & \hat{e}_2^{\prime} +\psi 
\hat{\overline{\eta}}_{\frac{3}{2}}^{\prime},  \nonu \\
f^{A=1}_{{\cal N}=4}(Z) \Big |_{\theta_3=\theta_4=0} 
& =&  \hat{\eta}_{\frac{3}{2}} + 
\psi \hat{e}_1 + \chi  \hat{A}_{0}  + \psi \chi 
\hat{\overline{\eta}}_{-\frac{1}{2}}, 
\label{twistorf} \\
f_{\dot{a}}^{{\cal N}=4}(Z) \Big |_{\theta_2=\theta_3=\theta_4=0} 
& =& \pa_{\dot{a}} \left( \hat{e}_2 + \psi 
\hat{\overline{\eta}}_{\frac{3}{2}} \right).  
\nonu
\eea
The helicity and $U(1)_R$ charge for $f^I(Z)$ are specified 
in section 3. Their assignments coincide with those in 
(\ref{twistorf}).  
The twistor field $\hat{\overline{\eta}}^
{\prime }_{\frac{3}{2}}$ corresponds to 
$\hat{\overline{\eta}}^
{\prime}_{\frac{3}{2}, A=1}$ of ${\cal N}=4$
twistor field and $\psi^{A=1}=\psi$.
After we put the $\theta_i=0$ where $i=2,3,4$,
all the higher order terms in $\psi^A$(quadratic and so on)
vanish.  
Now we move on the next twistor field description.
Similarly, the $\hat{\eta}_{\frac{3}{2}}$ corresponds to 
$\hat{\eta}_{\frac{3}{2}}^{A=1}$ of ${\cal N}=4$
twistor field.
$\hat{A}_{0}$ multiplied by 
two $\la$'s is equal to $\hat{V}^{B=2}_{1, A=1}$ of ${\cal N}=4$ 
twistor field(Note that the $\chi$ is given by $\la_1 \la_1 \psi^2$
at $\theta_3=\theta_4=0$ from previous footnote \ref{chifoot}).
The quantity $\hat{\overline{\eta}}_{-\frac{1}{2}}$ multiplied 
by two $\la$'s corresponds to 
 $\hat{\overline{\eta}}_{\frac{1}{2}, B=2}$
of ${\cal N}=4$ twistor field.  
Using the relation $\overline{\xi}_{[AB]}^B=0$ \cite{BW}, 
there is no 
such contribution from $(\psi^2)^{[BC]}
\hat{\overline{\xi}}^A_{\frac{1}{2}[BC]}$.
Of course, $\hat{e}_1$ stands for the same field as ${\cal N}=4$
case.
Finally $\hat{\overline{\eta}}_{\frac{3}{2}}$ corresponds to 
$\hat{\overline{\eta}}_{\frac{3}{2}, A=1}$ of ${\cal N}=4$
twistor field.

\section{Concluding remarks}
\setcounter{equation}{0}

\indent

In this paper, 
the spectrum of massless fields in spacetime described by 
the twistor fields are summarized by two equations 
(\ref{Ftwistor}) and 
(\ref{Gtwistor}).
By realizing the relation (\ref{abc})
between ${\cal N}=4$ chiral superfield and ${\cal N}=1$
chiral superfield,
the physical states of ${\cal N}=1$ conformal supergravity
are contained in the twistor superfields 
and they are summarized by the equations
(\ref{twistoradot}), 
(\ref{twistorA}), (\ref{twistora}) and (\ref{twistorf}).
We also put some results  and present their relations in both sides
in the Table 1.

It would be interesting to 
see 
the physical states of ${\cal N}=2$ conformal
supergravity from ${\cal N}=4$ superspace approach in terms of
the spacetime fields described by the twistor 
superfields.
For ${\cal N}=2$ conformal supergravity, 
the theory can be described off-shell by 
a chiral field strength ${\cal W}_{ab}^{AB}(x,\theta_a^A)$ 
where $A,B=1,2$
and the field contents \cite{BDD} for this theory have the extra
one antisymmetric tensor field, two spinors and three $SU(2)$
gauge fields as well as those in ${\cal N}=1$ conformal 
supergravity.
One can think of topological B-model of 
${\bf WCP}^{3|4}(1,1,1,1|1,1,p,q)$ where $p+q=2, p\neq 1, q\neq 1 
$ for 
the condition for Calabi-Yau 
supermanifold.
For example, when $p=-1$ and $q=3$, 
this is the projective space with four bosonic homogeneous 
coordinates $Z^I, I=1, \cdots,4$ of weights one, and four 
fermionic homogeneous coordinates $\psi, \chi, \al, \beta$
of weights one, one, minus one, and three. 
The homogeneous coordinates
have the following equivalence relation
$(Z^I, \psi, \chi, \al, \beta) \approx (t Z^I, t \psi, t \chi, 
t^{-1} \al, t^3 \beta)$ for $t$ is an element 
of ${\bf C}^{\ast}$.      
This supermanifold should admit ${\cal N}=2$ superconformal 
symmetry $SU(2,2|2)$ acting on $Z^I, \psi$ and $\chi$. 

\vspace{1cm}
\centerline{\bf Acknowledgments}
\indent

I would like to thank N. Berkovits for intensive discussions on 
his paper, and
A.D. Popov, W. Siegel,
P. Svrcek and 
E. Witten for discussions. 
Also I would like to thank the participants of IAS theory group 
meetings for getting some ideas during this year.
This research was supported by a grant in aid from the 
Monell Foundation through Institute for Advanced Study, 
by SBS Foundation, 
and by Korea Research Foundation Grant (KRF-2002-015-CS0006).

\end{document}